\documentclass[%
 reprint,
 amsmath,amssymb,
 aps
]{revtex4-1}
\usepackage{multirow}
\usepackage{graphicx}
\usepackage{dcolumn}
\usepackage{bm}
\usepackage{amsmath}
\usepackage{amssymb}
\usepackage{mathrsfs}
\usepackage{amsfonts}
\usepackage{color, soul}

\let\jnfont=\rm
\def\NPB#1,{{\jnfont Nucl.\ Phys.\ B }{\bf #1},}
\def\PLB#1,{{\jnfont Phys.\ Lett.\ B }{\bf #1},}
\def\EPJC#1,{{\jnfont Eur.\ Phys.\ Jour.\ C }{\bf #1},}
\def\PRD#1,{{\jnfont Phys.\ Rev.\ D }{\bf #1},}
\def\PRL#1,{{\jnfont Phys.\ Rev.\ Lett.\ }{\bf #1},}
\def\MPLA#1,{{\jnfont Mod.\ Phys.\ Lett.\ A }{\bf #1},}
\def\JPG#1,{{\jnfont J.\ Phys.\ G}{\bf #1},}
\def\CTP#1,{{\jnfont Commun.\ Theor.\ Phys.\ }{\bf #1},}
\def\ZPC#1,{{\jnfont Z.\ Phys.\ C }{\bf #1},}
\def\JHEP#1,{{\jnfont JHEP \ }{\bf #1},}
\def\Rv{\not{\hbox{\kern-1pt $R$}}}
\def\p{\not{\hbox{\kern-3pt $p$}}}

\newcommand{\lsim}{\mathrel{\lower4pt\hbox{$\sim$}}\hskip-10.5pt\raise1.6pt\hbox{$<$}\;}
\newcommand{\gsim}{\mathrel{\lower4pt\hbox{$\sim$}}\hskip-10.5pt\raise1.6pt\hbox{$>$}\;}

\newcommand{\beq}{\begin{eqnarray}}
\newcommand{\eeq}{\end{eqnarray}}
\newcommand{\bpmatrix}{\begin{pmatrix}}
\newcommand{\epmatrix}{\end{pmatrix}}

\newcommand{\ba}{\begin{array}}
\newcommand{\ea}{\end{array}}

\newcommand{\be}{\begin{equation}}
\newcommand{\ee}{\end{equation}}

\begin{document}

\preprint{ARXIV/1310.XXXX}

\title{ATLAS Z-peaked excess in the MSSM with a light sbottom or stop}

\author{Archil Kobakhidze$^1$}
\author{Ning Liu$^2$}
\author{Lei Wu$^1$}
\author{Jin Min Yang$^3$}
\affiliation{
$^1$ ARC Centre of Excellence for Particle Physics at the Terascale, School of Physics, The University of Sydney, NSW 2006, Australia\\
$^2$ Institute of Theoretical Physics, Henan Normal University, Xinxiang 453007, China\\
$^3$ State Key Laboratory of Theoretical Physics, Beijing 100190, China}%

\date{\today}

\begin{abstract}

We attempt to explain the recent ATLAS 3$\sigma$ excess of dilepton events with an invariant mass near $Z$ peak through the gluino-mediated sbottom production in a simplified scenario inspired by the Minimal Supersymmetric Standard Model (MSSM). The additional $Z$ bosons can be produced through the cascade decay chain $\tilde{g} \to b \tilde{b}^{\dagger}_{1} \to b \bar{b} \tilde{\chi}^{0}_{2,3} \to b \bar{b} \tilde{\chi}^{0}_1 Z$, in which $\tilde b_1$ is the right-handed sbottom, $\tilde\chi_1^0$ is the bino-like lightest supersymmetric particle (LSP) and $\tilde\chi_{2,3}^0$ are two nearly degenerate higgsino-like next-to LSPs (NLSPs). Taking into account the constraint from the LHC search for the gluino-mediated sbottom production in the final states with the missing transverse energy and at least three $b$-jets, we find that the ATLAS on-$Z$ excess can only be marginally explained at $2\sigma$ level. We also note that within the scenario where the gluino predominantly decays to the right-handed stop instead of the sbottom, the excess can hardly be explained, since the $Z$-boson is not produced in the stop dominant decay channel, $\tilde{t}_1 \to b \tilde{\chi}^+_1$.

\end{abstract}
\maketitle


\section{\label{sec:level1}Introduction}
After the discovery of a 125 GeV SM-like Higgs boson \cite{atlas,cms}, the focus of the LHC program has shifted to the searches for the new particles beyond the Standard Model (SM). Among many promising new physics models, the MSSM has been extensively investigated by theorists and experimentalists. At the LHC, the typical signature for the supersymmetry is the multi-jet plus missing transverse energy in the final states. Other topologies involving leptons, photons and heavy quarks have been explored in various simplified models as well. However, up to now, no significant excesses ($\sim 5\sigma$) beyond the SM were observed at the LHC.

Recently, the ATLAS collaboration has reported the $3\sigma$ excess in the on-$Z$ region in the search for the same-flavour opposite-sign dilepton pair, jets, and large transverse missing energy \cite{excess}. Including both of electron and muon pairs, the ATLAS collaboration observed 29 events after the selection cuts, while the expected SM background events are $10.6 \pm 3.2$. Although such an excess is currently not confirmed by the CMS analysis \cite{cms-dilepton}, it is interesting to study it in various new physics models. In Ref. \cite{spain}, the authors argued that the ATLAS on-$Z$ excess can be explained in the general gauge-mediation (GGM) framework. But later, in Ref. \cite{allanach}, the authors claimed that such a model can not explain the excess due to the LHC constraints. In Ref. \cite{ellwanger}, considering the constraints from the LHC searches for the multi-jet and the missing transverse energy \cite{atlas-multijets}, the author proposed an explanation through gluino pair production in the Next-to-Minimal Supersymmetric Standard Model (NMSSM). Besides SUSY interpretations, a heavy gluon decay to the vector-like quarks in the composite Higgs theories was suggested to explain the ATLAS on-$Z$ excess \cite{composite}.

In many popular supersymmetric grand unification models, the first-two generation squarks appear to be heavier than the third generation squarks, due to the different renormalization effects related to the Yukawa coupling contributions. In particular, a right-handed sbottom can be light in some models \cite{bartl}. A heavy stop sector is usually required by the observed 125 GeV Higgs boson in the MSSM, however, the large mixing effect can also achieve the correct Higgs mass and lead to a relatively light stop. Therefore, if sufficiently light, the third generation squarks could be copiously produced through the gluino decay. On the other hand, the gluino-mediated the third generation squarks productions have been searches for at the LHC. The null results give the strong bounds on the stop/sbottom and the gluino mass \cite{atlas-gluino,cms-gluino}.

In this work, we focus on the gluino-mediated sbottom production with $\tilde{b}_{1} \to b \tilde{\chi}^{0}_{2,3} \to b \tilde{\chi}^{0}_1 Z$ in a simplified scenario inspired by the MSSM and examine its interpretation for the ATLAS $Z$-peaked excess reported under the LHC constraints. We also study the feasibility of the explanation of the gluino-mediated stop production for the excess. The structure of this paper is organized as follows. In Section \ref{section2}, we will set up our calculations and present the numerical results. Finally, we draw our conclusions in Section \ref{section3}.

\section{\label{section2}Calculations and discussions}
In order to reduce the SM background events, the following requirements are applied in the ATLAS searches for the final states containing a same-flavour opposite-sign lepton pair, jets and large missing transverse momentum.
\begin{itemize}
\item Events are required to have at least two same-flavour opposite-sign (SF-OS) leptons. If more than two signal leptons are in a event, the two with the hardest $p_T$ are selected. The leading lepton must have $p_T(\ell_1) > 25$ GeV and the sub-leading lepton $p_T(\ell_2)$ can be as low as 10 GeV. The invariant mass of these two leptons has to be in the $Z$ boson mass window 81 $< m_{\ell^+\ell^-} <$ 101 GeV.
\item All events are further required to contain at least two hard jets with $p_T > 35$ GeV  and $|\eta| < 2.5$, to have missing transverse energy  $E^{miss}_T> 225$ GeV  and $H_T > 600$ GeV, where $H_T$ is the $p_T$ sum over all the jets with  $p_T > 35$ GeV  and $|\eta| < 2.5$ and the two leading leptons: $H_T = \sum_i p_{T}^{jet,i} + p_{T}^{lepton,1} + p_{T}^{lepton,2}$.
\item Besides, the azimuthal angle between each of the two leading jets and $E^{miss}_T$ has to be larger than 0.4.
\end{itemize}
From the above selection criteria, we can infer that a strong production process is more favored than an electroweak process to explain the ATLAS observed on-$Z$ excess due to the requirement of the multiplicity of jets and a hard $H_T$ cut \cite{spain}. So, we will focus on the gluino production in the following.

\begin{figure}[ht]
\centering
\includegraphics[width=3in,height=2in]{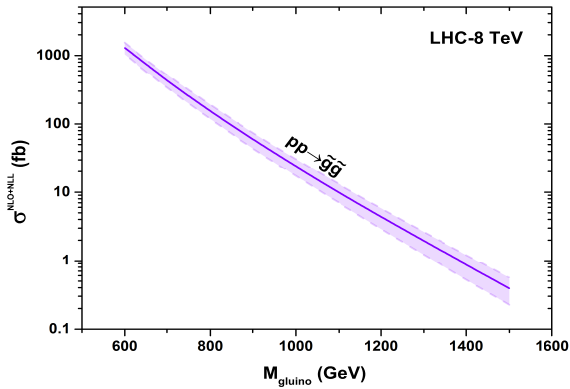}
\caption{The NLL+NLO cross section of gluino pair production for decoupled squarks at 8 TeV LHC. The purple band corresponds to $1\sigma$ theoretical uncertainty.}
\label{cx}
\end{figure}
In Fig.\ref{cx}, we calculate the NLO+NLL cross section of the gluino pair production at 8 TeV LHC by using \textsf{NLL-fast} package \cite{nll-fast} with the CTEQ6.6M PDFs \cite{cteq6}. We can see that the cross section of the gluino pair $\sigma_{\tilde{g}\tilde{g}}$ can still reach tens of fb when the gluino mass $m_{\tilde{g}} \lesssim 1$ TeV. If there is no branching ratio suppression in the gluino cascade decay chain, assuming $m_{\tilde{g}}=1$ TeV and the signal events keeping efficiency $\epsilon \sim 50\%$, one can obtain about 25 on-$Z$ events at 8 TeV LHC with the luminosity of ${\cal L}=20.3$ fb$^{-1}$. Any reduction in the branching ratio or in the event keeping efficiency will require the gluino to be light to enhance the total cross section. However, a light gluino (with the mass less than 1 TeV) usually suffers from the strong LHC constraints. Thus, the most economic way to explain the ATLAS on-$Z$ excess is to make the branching ratio of the gluino cascade decay with a $Z$ boson in the final states as large as possible.

\begin{figure}[ht]
\centering
\includegraphics[width=3in,height=1in]{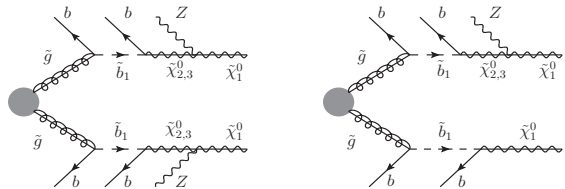}
\caption{Gluino-mediated sbottom production contributing to on-$Z$ signal.}
\label{feyn}
\end{figure}
In Fig.~\ref{feyn}, we show the diagram of the gluino decay chain that is relevant for our study. If $m_{\tilde{b}_1}<m_{\tilde{g}}$, the produced gluino will dominantly decay to the sbottom. However, the sbottom decay depends on the nature of sbottom and the electroweakinos. The interaction between sbottom and neutralinos is given by \cite{mssm-feynrules}
\begin{eqnarray}
{\cal L}_{\tilde{b}\bar{b}\tilde{\chi}^0_i} &=& \tilde{b}_L \bar{b} \left[ y_b N_{i3} P_L + \sqrt{2} (\frac{g}{2} N_{i2} - \frac{g'}{6} N_{i1})P_R \right] \tilde{\chi}^0_i  \nonumber \\
&& + \tilde{b}_R \bar{b} \left[ -\frac{\sqrt{2}}{3}g' N_{i1} P_L + y_b N_{i3} P_R \right] \tilde{\chi}^0_i + h.c.,~
\label{vertex}
\end{eqnarray}
where $P_{L/R}=(1\mp\gamma_5)/2$ and $y_b=\sqrt{2}m_b/(v\cos\beta)$. The mixing matrix $N_{ij}$ of the neutralino sector are defined in \cite{mssm-feynrules}. For $\tilde{\chi}^0_{2,3}$ being higgsino-like neutralino and $\tilde{\chi}^0_{1}$ being bino-like neutralino, we can see that the right-handed sbottom can have the sizable decay branching ratios of $\tilde{b}_1 \to b\tilde{\chi}^0_{2,3}$ in comparison with the case of $\tilde{\chi}^0_{2,3}$ being wino-like neutralinos \cite{wu-1}. In particular, when $\tilde{b}_1 \to t\tilde{\chi}^-_{1}$ is not allowed by the phase space, the decay of $\tilde{b}_1 \to b\tilde{\chi}^0_{2,3}$ will be dominant. Then, the higgsino-like neutralinos $\tilde{\chi}^0_{2,3}$ can exclusively decay to the $Z$ boson, if $m_Z < m_{\tilde{\chi}^{0}_{2,3}} - m_{\tilde{\chi}^{0}_{1}} < m_h$. So, we study a simplified scenario of the MSSM, which consists of a heavy gluino ($\tilde{g}$), a right-handed sbottom ($\tilde{b}_{1}$), a bino-like LSP ($\tilde\chi_{1}^0$), two neutral ($\tilde\chi_{2,3}^0$) and one charged ($\tilde\chi_{1}^\pm$) nearly degenerate higgsino-like NLSPs. Other sparticles are assumed to be heavy and are decoupled for simplicity. In Tab.~\ref{bp1}, we use a MSSM benchmark point to illustrate our arguments, where the \textsf{Suspect} \cite{suspect} and \textsf{SUSY-HIT} \cite{susyhit} are used to calculate the masses, the couplings and the branching ratios. All the soft SUSY breaking mass parameters are defined at the weak scale.

\begin{table}[ht]
\begin{center}
\begin{tabular}{|c|c|c|c|}
\hline
 $M_1$ &$M_2$ &$M_3$ &$\mu$  \\
\hline
550 &2000 &840  &650   \\
\hline\hline
$\tan\beta$ &$m_A$ &$M_{\tilde{q}^{1,2}_{L,R}}$ &$M_{\tilde{Q}_{3L},\tilde{t}_R}$ \\
\hline
50  &1000 &5000 &2000 \\
\hline\hline
$A_{b,t}$ &$M_{\tilde{b}_R}$ &$m_{\tilde{g}}$ &$m_{\tilde{b}_1}$ \\
\hline
0 &800 &840 &799.51 \\
\hline\hline
$m_{\tilde{\chi}^0_1}$ &$m_{\tilde{\chi}^0_2}$ &$m_{\tilde{\chi}^0_3}$ &$m_{\tilde{\chi}^\pm_1}$  \\
\hline
541.23 &651.88 &657.23 &648.77 \\
\hline \hline
$\sigma({\tilde{g}\tilde{g}})$ &Br(${\tilde{g}\to \tilde{b}_1\bar{b}}$) &Br(${\tilde{g}\to \tilde{b}^\dagger_1 b}$) &Br(${\tilde{b}_1 \to b \tilde{\chi}^0_{1}}$)   \\
\hline
104 fb &50\% &50\% &24.88\% \\
\hline\hline
Br(${\tilde{b}_1 \to b \tilde{\chi}^0_{2}}$) &Br(${\tilde{b}_1 \to b \tilde{\chi}^0_{3}}$)  &Br(${\tilde{\chi}^0_2 \to Z \tilde{\chi}^0_{1}}$) &Br(${\tilde{\chi}^0_3 \to Z \tilde{\chi}^0_{1}}$) \\
\hline
40.31\% &34.81\% &100\% & 100\% \\
\hline
\end{tabular}
\caption{ A benchmark point for the gluino-mediated sbottom production. Here the mass parameters $M_{1,2,3}$ (gaugino masses), $\mu$ (higgsino mass), $M_{\tilde{q}^{1,2}_{L,R}}$ (the soft breaking masses for the first two generation squarks), $m_A$ (the CP-odd Higgs boson mass), $M_{\tilde{Q}_{3L},\tilde{t}_R,\tilde{b}_R}$ (the soft breaking masses for the third generation squarks) and $A_{t,b}$ (the third generation trilinear couplings) are in unit of GeV. $\tan\beta$ is the ratio of the vacuum expectation values of the two-Higgs doublet fields in the MSSM. The gluino pair cross section is calculated up to NLL+NLO for LHC-8 TeV. \label{bp1}}
\end{center}
\end{table}

Next, we examine the LHC constraints on our scenario and investigate the ATLAS on-$Z$ excess in the allowed parameter space region. In general, one can vary the mass splitting $m_{\tilde{b}_{1}} - m_{\tilde{\chi}^{\pm}_{1}}$ in the range of $(m_b, m_t)$ and $m_{\tilde{\chi}^{0}_{2,3}} - m_{\tilde{\chi}^{0}_{1}}$ in the range of $(m_Z, m_h)$, respectively. However, the smaller mass differences become, the fewer on-$Z$ events are kept, due to the requirement of a large value of $H_T$ in the ATLAS analysis. Thus, we assume $m_{\tilde{b}_{1}} - m_{\tilde{\chi}^{\pm}_{1}}=150$ GeV and $m_{\tilde{\chi}^{0}_{2,3}} - m_{\tilde{\chi}^{0}_{1}}=100$ GeV in the gluino decay chain. Finally, only the gluino mass and the right-handed sbottom mass are the free parameters in the following scan of our parameter space. Other irrelevant sparticles are assumed to be heavy, which include the sleptons, the first two generations of squarks, the stops, the winos, the heavy CP-even Higgs boson and the CP-odd Higgs boson.

To study the LHC direct search constraints on our parameter space, we use the package \textsf{CheckMATE-1.2.1} \cite{checkmate-1} to recast the ATLAS analysis of the gluino-mediated sbottom production \cite{atlas-gluino} and the search for the multi-jet with transverse missing energy events \cite{atlas-multijets}. Besides, we reinterpret the CMS search for the multi-jet events \cite{cms-multijet} to examine its constraint on the parameter space by using the package \textsf{MadAnalysis 5-1.1.12} \cite{ma5}. Both of the packages use the tool \textsf{FastJet} \cite{checkmate-2}, where the anti-$k_t$ algorithm is chosen for the jet clustering \cite{checkmate-3}. The validations of these analyses have been done by the two groups. The parton level signal events are generated by the package \textsf{MadGraph5} \cite{mad5} and are showered and hadronized by the package \textsf{PYTHIA} \cite{pythia}. The detector simulation effects are included with the tuned package \textsf{Delphes} \cite{delphes}, which is contained in \textsf{CheckMATE-1.2.1} and \textsf{MadAnalysis 5-1.1.12}. Finally, we define the ratio $r = max(N_{S,i}/S^{95\%}_{obs,i})$ for each experimental search, where $N_{S,i}$ is the number of the signal events for the $i$-th signal region and $S^{95\%}_{obs,i}$ is the corresponding observed 95\% C.L. upper limit. The max is over all the signal regions for each search. If $r > 1$, we conclude that such a point is excluded at 95\% C.L..

\begin{figure}[ht]
\centering
\includegraphics[width=3in,height=3in]{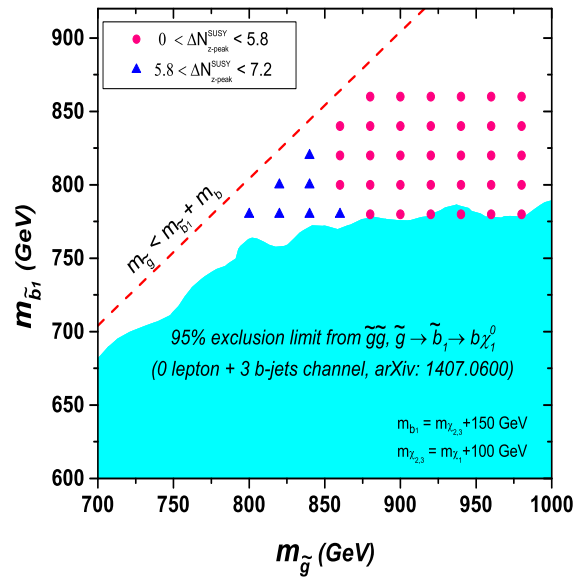}
\caption{The on-$Z$ signal events from the gluino-mediated sbottom production. The blue triangles and pink bullets represent the events number of the on-$Z$ signals that can explain the ATLAS excess in and outside 2$\sigma$ range, respectively. The cyan region is excluded by the ATLAS search for gluino production in final states with missing transverse energy and at least three $b$-jets \cite{atlas-gluino}.}
\label{zpeak}
\end{figure}
In Fig.~\ref{zpeak}, we present the on-$Z$ signal events from the gluino-mediated sbottom production on the plane of $m_{\tilde{b}_1}$ and $m_{\tilde{g}}$. The blue triangles and pink bullets represent the number of the on-$Z$ events that can explain the ATLAS excess in and outside 2$\sigma$ range, respectively. The cyan region is excluded by the ATLAS search for the gluino-mediated sbottom production in the final states with the missing transverse energy and at least three $b$-jets \cite{atlas-gluino}, which is found to be the most stringent bound on our scenario. From the Fig.~\ref{zpeak}, we can see that when $m_{\tilde{g}}$ is lighter than 800 GeV, the mass difference $m_{\tilde{g}}-m_{\tilde{b}_1}$ should be less than about 20 GeV to avoid the constraint. However, such a small splitting can not produce the large value of $H_T$ that is required by the ATLAS analysis of the on-$Z$ signal. If $m_{\tilde{g}}$ becomes heavy, the cross section of the gluino pair production will reduce rapidly. So, we find that only a small parameter region with $800~{\rm GeV} \lesssim m_{\tilde{g}} \lesssim 860~{\rm GeV}$ and $760~{\rm GeV} \lesssim m_{\tilde{b}_1} \lesssim 820~{\rm GeV}$ can marginally explain the ATLAS on-$Z$ excess at $2\sigma$ level.

We also have considered a similar scenario where the gluino dominantly decays to the right-handed stop with the sbottom being decoupled. In this case, the $Z$ bosons can be produced through the decay chain $\tilde{t}_1 \to t \tilde{\chi}^0_{2,3} \to t Z \tilde{\chi}^0_1$. But the branching ratios of $\tilde{t}_1 \to t \tilde{\chi}^0_{2,3}$ are usually less than $50\%$ since $\tilde{t}_1 \to b \tilde{\chi}^+_1$ is the dominant decay mode of the right-handed stop \cite{wu-2}. Besides, such a scenario are strongly constrained by the LHC searches for the multi-jet events and the gluino-mediated stop production. So it is hard to explain the ATLAS on-$Z$ excess through the gluino-mediated stop production.

In the above discussions, only the light right-handed sbottom or stop is involved in the gluino cascade decay process. If both of them are light enough, the gluino will decay to $b\tilde{b}^\dagger_1$ and $t\tilde{t}^\dagger_1$. This will lead to the suppression of branching ratios of $\tilde{g} \to \tilde{b}_1 \to b \tilde{\chi}^0_{2,3}$ and fewer on-$Z$ signal events than the above two cases. So, we can conclude that such a light sbotttom and stop scenario can not give a better explanation than our studied sbottom scenario.

Given the current excess is only reported by the ATLAS, with more data in LHC run-2, both ATLAS and CMS analyses will be able to further confirm this excess if it is indeed a signal of new physics beyond the SM. On the other hand, no matter this anomaly will persist or not, a compressed gluino-mediated sbottom production is an interesting scenario to be explored at the future LHC \cite{fuks,datta}.

\section{\label{section3}conclusion}

In this work, we study the possible explanation of the recent ATLAS 3$\sigma$ on-$Z$ excess through the gluino-mediated sbottom production in a simplified scenario inspired by the MSSM, where the spectrum consists of a gluino, a right-handed sbottom, two neutral and one charged nearly degenerate higgsino-like neutralino NLSPs and a bino-like neutralino LSP. Under the LHC constraints, in particular the ATLAS search for gluino-mediated sbottom production in final states with the missing transverse energy and at least three $b$-jets, we find that the ATLAS on-$Z$ excess can only be marginally explained at $2\sigma$ level in a narrow allowed parameter space region, where the mass splittings between the gluino and the sbottom are less than about 100 GeV. Besides, we note that if the gluino decays to right-handed stop instead of sbottom, it is hard to explain the excess because the stop will largely decay to $b \tilde{\chi}^+_1$ without $Z$-boson being produced.

\subparagraph{Acknowledgments} We thank the early stage collaborations with Aldo Saavedra. This work was supported by the Australian Research Council, by the National Natural Science Foundation of China (NNSFC) under grants No. 11275057, 11305049, 11405047, by Specialized Research Fund for the Doctoral Program of Higher Education under Grant No. 20134104120002.

\end{document}